# Blue Emission in Proteins

Sohini Sarkar, Abhigyan Sengupta, Partha Hazra,* and Pankaj Mandal*

*Department of Chemistry, Indian Institute of Science Education and Research, Pune 411008, India*

Abstract: Recent literatures reported blue-green emission from amyloid fibril as exclusive signature of fibril formation. This unusual visible luminescence is regularly used to monitor fibril growth. Blue-green emission has also been observed in crystalline protein and in solution. However, the origin of this emission is not known exactly. Our spectroscopic study of serum proteins reveals that the blue-green emission is a property of protein monomer. Evidences suggest that semiconductor-like band structure of proteins with the optical band-gap in the visible region is possibly the origin of this phenomenon. We show here that the band structure of proteins is primarily the result of electron delocalization through the peptide chain, rather than through the hydrogen bond network in secondary structure.

Unrestrained protein association may result in deleterious consequences like amyloid plaque deposition culminating in numerous neurodegenerative disorders such as Alzheimer's and Parkinson's diseases.[1,2] Therefore, extensive spectroscopic studies have been carried out for the development and quantitative application of fluorescent assays to monitor the proteins self-assembly processes.[3] Recently, it has been reported that the amyloid fibrils formed from protein as well as small peptides exhibit luminescence in blue and green region, although the peptide or protein is devoid of chromophores which can emit in the visible region.[4-6] Guptasarma and co-workers reported about the blue fluorescence in protein crystal as well as in solution of γ-B-crystallin protein for the first time, and they attributed this unusual intrinsic fluorescence to the delocalization of electron through hydrogen bonding.[7,8] Smith *et al.* suggested that the blue emission is related to the formation of extensive J-aggregates that arise from π-stacking of aromatic moieties. However, Chan *et al.* argued that the intrinsic fluorescence observed in human peptides amyloid-β(1-42), lysozyme and Tau is independent of the presence of aromatic side chain residues; rather electronic delocalization via hydrogen bonds in β-sheet structure can cause this phenomenon.[9,10] This view has been supported by Sharpe *et al*.[11] Anand and Mukherjee ascribed the visible fluorescence of amyloid fibril obtained from peptide to the excitonic transition.[4] Very recently, Chatterjee *et al*. reported concentration dependent emission study of Bovine Serum Albumin (BSA), and attributed the blue fluorescence to protein aggregation.[12] Evidently, the origin of this 'unusual visible luminescence' is highly debated. Yet it is believed that blue-green fluorescence is a generic property of amyloid structures, and efforts are made to use this fluorescence to monitor the fibril growth.[3]

Herein, we report a systematic study on intrinsic fluorescence property of serum proteins, copiously present in blood serum of vertebrates, to find the origin of this blue fluorescence. Our study proves that blue-green fluorescence of protein can be observed without formation of amyloid fibril or any other form of aggregation and is, indeed, a property of monomeric protein unit. The correlation between our optical study and the existing conductivity data of proteins shows that the semiconductor band structure of proteins with a band-gap in the visible region is the possible origin for this visible luminescence. Our finding also suggests that electron delocalization through the peptide chain, rather than through hydrogen bond network, plays a more significant role in determining the band structure of protein.

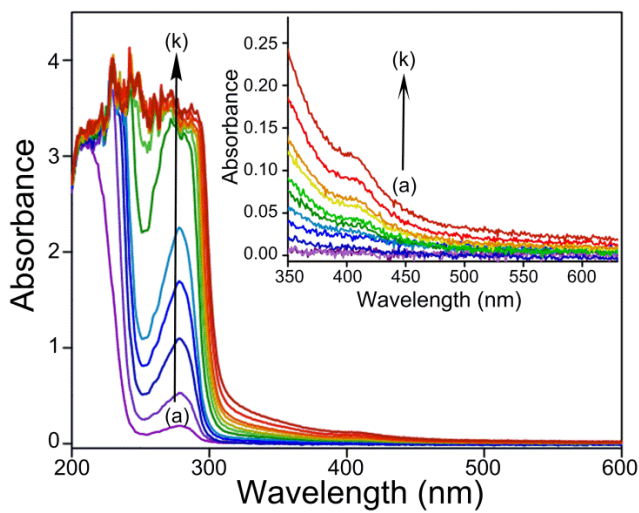

**Figure 1.** UV-visible absorption spectra of different concentration of BSA solution where (a) to (k) represent 4.8, 12, 36, 48, 96, 144, 192, 240, 288, 384 and 480 μM of BSA. The inset shows an expanded view of the hump at ~410 nm, visible at higher concentration of protein.

In our attempt to systematically analyze the origin of blue fluorescence of proteins, we initially recorded the concentration dependent absorption spectra of bovine serum albumin (BSA) as depicted in Figure 1. Tryptophan (Trp 134 and Trp 212), being the intrinsic chromophore in BSA, shows a prominent absorption at 278 nm.[13] Interestingly, at 50 μM, a hump at ~410 nm becomes visible and grows stronger with higher BSA concentrations with the absorption tail extending till 550 nm. This peeping hump at ~410 nm might be an outcome of structurally reorganized low energy ground state configuration of weakly bound protein assembly/aggregation at higher concentration[14] or low energy ground state of the protein monomer unit with very small oscillator strength.

It is well established that Trp is the major origin of BSA fluorescence and emits at ~350 nm (Figure S2 in Supporting Information (SI)).[13] Serum proteins, being devoid of any chromophore which can absorb visible light, are not expected to fluoresce in visible or near IR region. Surprisingly, we observed that serum proteins exhibit concentration and excitation wavelength dependent fluorescence in the visible region. Similar results have also been reported earlier.[8, 12] On exciting at 360 nm, an emission band centered at 450 nm is observed at protein concentrations as low as 5 μM. Emission at 450 nm at lower BSA concentration (~5 μM) is extremely weak but certainly present, and increases monotonically with increasing concentration (Figure 2). Similar result is also observed in case of human serum albumin (HSA) and lysozyme (Figure S4 in SI).

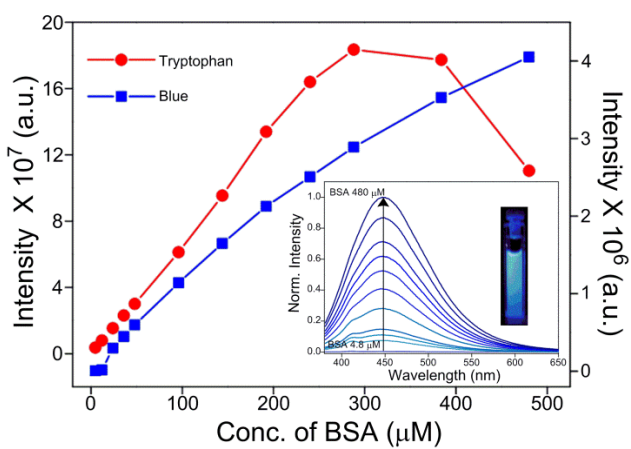

**Figure 2.** Variation of fluorescence peak intensity (corrected for inner filter effect using dual path method; see SI for details) with increasing concentration of BSA. Excitation wavelengths for Trp emission (red circle) and blue emission (blue square) were 295 nm and 360 nm respectively. **Inset:** Blue emission spectra at different concentration of BSA and a real photograph of the cuvette having 480 μM BSA under UV illumination.

It is interesting to compare the concentration dependent variation of two different emission peaks (350 nm and 450 nm), as shown in Figure 2 (see also Figure S2 in SI). The Trp emission at 350 nm increases linearly with BSA concentration up to ~300 μM and start decreasing at higher concentration, whereas the 450 nm peak keeps growing monotonically within the concentration range of present study. A very similar quenching of Trp emission with the increase in concentration of macromolecular crowder has been reported recently.[15] The authors concluded that this was mostly a case of static quenching where the presence of crowder molecules induces conformational change in the protein molecule. As a result the Trp moiety becomes surrounded by protonated acidic groups such as Lysine (Lys), Histidine (His), Aspartic Acid (Asp) and Glutamic acid (Glu), which quench the fluorescence of Trp by increasing the hydrophilic nature of the immediate surroundings of Trp.[15] So, at higher concentration protein molecules may experience macromolecular crowding but it does not really affect the blue emission of protein.

Another intriguing finding is the excitation wavelength dependent blue emission profile of BSA (Figure S6 in SI). However, the Trp emission at 350 nm does not shift with the change in excitation wavelength. The excitation-wavelength dependent emission maxima shift, as observed for BSA fluorescence, is regularly observed for fluorophore in proteins, vesicles, membranes and micelles, and this phenomenon is known as Red Edge Excitation Shift (REES). This phenomenon occurs when the solvent relaxation time of the excited state is comparable or longer compared to the fluorescence life time.[16,]

[17] We found that the fluorescence lifetime of the blue emission is ~1.2 ns and does not really depend much on the concentration of BSA (Table S1 in SI).

The recent report claims that the blue emission of BSA is a result of aggregation at higher concentration.[12] However, we observed the blue emission even at concentration as low as 5 μM. At this concentration the volume fraction of the protein is ~0.01%. It seems quite unlikely that the protein molecules should undergo aggregation in such a condition. So, we conducted several experiments to find out if proteins aggregate or not at higher concentration; and if they do, whether the aggregates are necessary requirement for blue emission or not. First of all, we determined the hydrodynamic radii of particles (protein monomer, oligomers or aggregates) present in BSA solutions of different concentrations using Dynamic Light Scattering (DLS) as shown in Figure 3A (see also Figure S8 in SI). In contrast to the DLS data of the recent report,[12] our study shows only one mean diameter (~8.5 nm), which is consistent with the effective geometry corresponding to monomeric hydrodynamic radius for BSA reported in earlier works, at all concentrations (10 μM to 250 μM) studied.[12, 18] We did not find any second peak proving the presence of higher aggregates. However, the width of the distribution increases slightly at higher concentration, indicating presence of oligomers. To validate our DLS finding, we conducted size exclusion chromatograpy (SEC) of protein solutions at two very different concentrations, 10 and 100 μM. SEC, also known as gel filtration, is a proven and regularly practiced method for separating bio-molecules on the basis of their sizes. Both solutions provided two distinct fractions, one with monomeric size (66.4 kD) and the other having higher order structures (dimer, trimer or tetramer) with molecular mass in the range of 130 - 270 kD at a ratio of 4 to 1 (Figure S9 in SI). The monomeric fraction having an effective concentration of 7 μM produced fluorescence spectrum, identical to that of parent solution, with emission peak at 450 nm when excited with 360 nm light (Figure S10 in SI). The above results clearly show that aggregation does not play any role towards the origin of the blue fluorescence of proteins.

To find out if the monomeric protein unit is capable of emitting blue light or the oligomeric structures are responsible for the observed fluorescence, we recorded fluorescence spectrum of protein molecule confined inside a reverse micelle (RM). Reverse micelle is a biphasic nano-sphere with a hydrophobic surface encapsulating a water pool in the core. The size of the water pool can be tuned by varying the water to surfactant molar ratio ($W_0$). We used bis-(2-ethylhexyl) sodium sulfosuccinate (AOT) RM and deliberately chose a core size of 5.9 nm ($W_0$ = 25), which would be just large enough to accommodate a single protein molecule (effective hydrodynamic radius ~ 3.5 nm) but not oligomers.[19, 20] This arrangement certainly excluded the possibility of forming larger aggregates. As shown in Figure 3B, the protein monomers (~12 μM) confined in the RM also emit blue light (peak emission ~450 nm) when excited at 360 nm. The nature of the fluorescence profile of BSA in RM matches quite well with that of BSA in bulk solution having similar concentration. However, Trp emission at 350 nm also shows REES along with regular REES as observed for the 450 nm emission, suggesting a possible change in conformation of protein in RM (Figure S11 in SI). This observation undoubtedly proves that the blue emission of the protein is an intrinsic property of the monomeric unit itself.

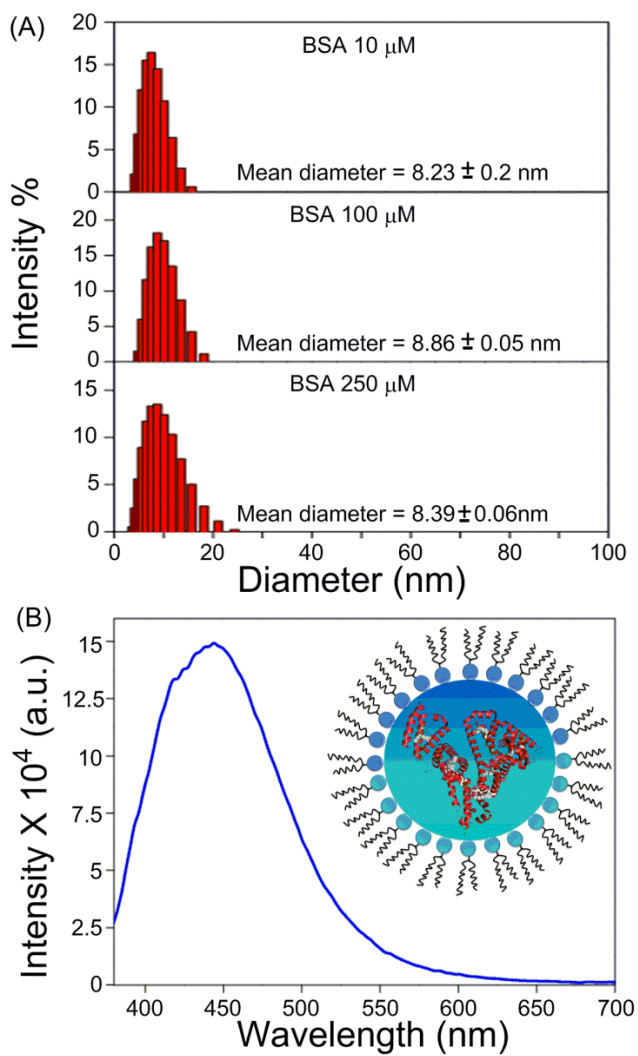

**Figure 3.** (A) Size distribution histogram from DLS measurements for 10, 100 and 250 µM BSA. (B) Fluorescence spectrum of BSA monomer confined within AOT reverse micelle with a core size of 5.9 nm. **Inset:** Schematic of a protein monomer confined inside RM.

Having established that the ubiquitous blue from fluorescence protein comes from the protein monomer itself and not due to aggregation, further we tried to disentomb the origin of unusual blue fluorescence. To find out if the secondary and tertiary structure of protein is responsible for this blue fluorescence or not, we studied the effect of denaturation on the fluorescence spectra of protein. Figure 4 displays the variation in normalized intensity of the blue emission (450 nm) as well as the Trp emission (350 nm) as a function of concentration of Guanidinium chloride (GdmCl), a well known denaturant,[21] added to 144 µM BSA solution. The Trp emission quenches much more aggressively compared to the quenching of blue emission on addition of denaturant. We also observed a prominent red-shift (~10 nm) of the Trp emission and a very small blue-shift of the blue emission at higher denaturant concentration (Figure S12 & S13 in SI). The pH dependent fluorescence study in the pH range of 4 to 12 shows that the intensity of the 450 nm emission remains virtually unaffected in the pH range of 4 to 10, but drops by ~25% at pH 12 (Figure S14 in SI). On the contrary, the Trp emission quenches gradually on increasing pH above 7, and dramatically drops by ~98 % at pH 12 suggesting a complete denaturation of the protein. The above results clearly suggest that the blue emission of protein exists even after denaturing the protein completely, discarding the possibility of secondary and tertiary structures to be responsible for the blue fluorescence.

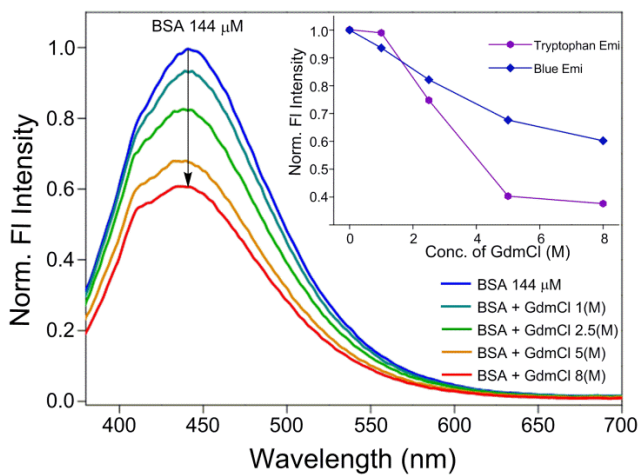

**Figure 4.** Normalized blue emission spectra (excitation wavelength is 360 nm) at different concentration of Guanidinium chloride (GdmCl) added to BSA solution (144 µM). **Inset:** Change in peak intensity of blue emission (450 nm) and Trp emission (350 nm) with increasing denaturant (GdmCl) concentration.

When we applied thermal energy (heat) for gradual denaturation of the protein, a striking observation led to exploring the possible origin of blue fluorescence of protein. As expected from the denaturation experiments mentioned above, temperature dependent fluorescence study of BSA solution shows that the blue fluorescence intensity decreases with increasing temperature (Figure S15 in SI). In addition we found a small gradual red-shift of the peak maxima on increasing the temperature from 10˚C to 60˚C as shown in Figure 5. Such a phenomenon is often observed for semiconducting materials.[22] Higher thermal energy increases the population at the higher edge of the conduction band, thus decreasing the effective band-gap of the semiconductor. Szent-Györgyi, for the first time in 1941, suggested that a protein molecule can have a band structure similar to a semiconductor, and it can become conducting once the activation (band-gap) energy is supplied.[23, 24] S. Baxter experimentally demonstrated that a protein (wool) behaved as a semiconductor.[25] Since then it has become a very active field of theoretical and experimental research that deal with conductivity of proteins and other bio-polymers. Early theories considered that the lone pair electrons and the π-electrons of the amide moiety can delocalize through the hydrogen bonds in secondary structure to produce (pseudo)bands of very closely spaced energy states, instead of distinct energy levels.[26-28] Figure 6 schematically represents possible origin of band structure of proteins as hypothesized by Evans and Gergely.[29] Protein molecule has many peptide units (583 amino acids for BSA) and is essentially a polymer. Each amino acid moiety will have very similar energy levels, but not same because of the perturbation from different side-chains and a slightly different immediate neighbour. So, instead of a single energy state, a band of energy consisting of $n$ numbers (for a poly-peptide with $n$ number of peptide moiety) of closely spaced energy levels will be present. This will produce a band structure very similar to one present in a semiconductor material. Similarly other polymeric systems are also capable of forming band structure. This band structure is well known and often mentioned in conducting polymers. In case of conducting polymer the conduction band is partially filled. For proteins or polypeptide chains the valence band is completely filled and the conduction band is completely empty. If the electrons are promoted to the conduction band by supplying the energy equal or more than the band-gap, the electron is free to move in the conduction band. Several theoretical studies have reported considerable mixing of sigma and pi orbitals in energy bands.[26, 30] Because of the significant contribution of sigma-orbitals into the conduction band, the electron in the conduction band can delocalize along the peptide chain in protein.[26-28] However, a debate regarding the relative contribution of two pathways of electron-delocalization mentioned above still persists.

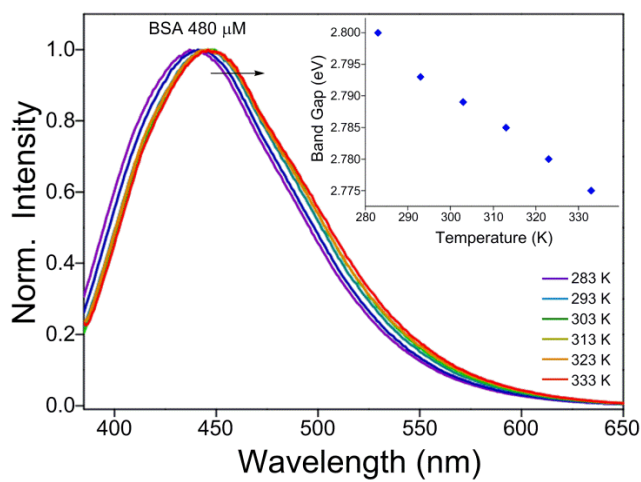

**Figure 5.** Temperature dependence of blue fluorescence in the range of 10°C to 60°C. **Inset:** Change in emission peak wavelength (in energy unit eV, denoted as "band gap") with temperature.

Eley and Spivey measured the conductivity of dry BSA and Lysozyme in 1960 and reported thermal band-gaps of 2.78eV (~446 nm) and 2.62 eV (~473 nm) respectively.[31] Comparing these thermal band-gap values with our optical studies, we strongly believe that the absorption peak at ~ 410 nm (Figure 1) is due to transition from valence band to conduction band, and the blue emission at 450 nm is the result of subsequent radiative recombination of electron and hole. The fact that even the denatured protein shows blue emission, having effectively same emission peak wavelength, (Figures 4; S14, & S15 in SI) suggests that the hydrogen-bond network in the secondary structure may not be of much significance in creating the band structure of protein. Instead, electron delocalization through peptide chain seems to play major role.

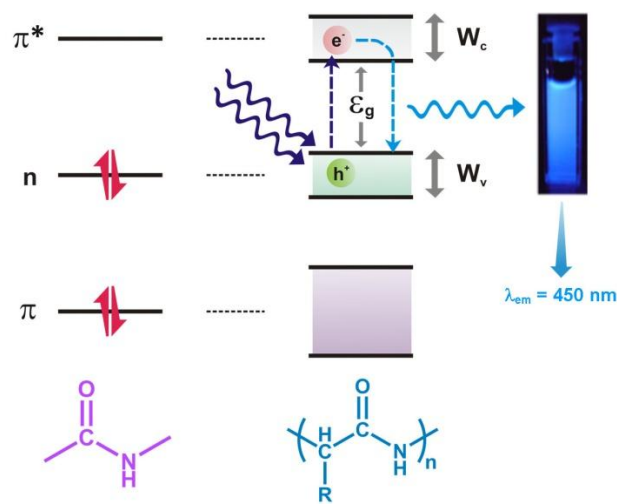

**Figure 6:** Schematic depicting the origin of band structure of proteins and corresponding blue fluorescence.

However, the small blue-shift of the emission maxima upon denaturation of the protein (Figure S13 in SI) may indicate a small possible contribution of the electron delocalization through the hydrogen bond network towards the band structure of the protein. Very similar optical band-gap (band-edge emission ~ 450 nm) of BSA (583 amino acid) and lysozyme (129 amino acid) indicates that even a shorter polypeptide chain (band forming unit) may be sufficient to produce an optical band-gap in the blue-green region. Many other proteins and polypeptides of different sizes also show thermal band-gap in the range of 1.5 to 3.5 eV (830 nm to 350 nm).[32] The decrease in intensity of the blue emission in denatured protein may be attributed to the unfolded structure of protein having a smaller number of semiconducting polypeptide unit per unit volume. In other words, the effective concentration of the band-forming units decreases on unfolding.

Several dendrimers with –NH and –OH groups in the repeating unit[33] and polysaccharides (having -OH group and -O-linkage) also emit similar unusual blue-green fluorescence even though these macromolecules don't have any traditional chromophore (Figure S16 in SI). Similar to polypeptide chain, these polymeric macromolecules also have groups with pi-electron and/or lone pair of electrons that can delocalize through the polymer chain creating similar semiconducting band structure with optical band-gap in the visible range.

In summary, we have established that the blue fluorescence is not a characteristic feature of amyloid fibril or aggregation, but a property of monomeric protein unit. Moreover, the blue fluorescence does not depend much on the secondary and tertiary structure of the proteins. We strongly believe that this unusual visible fluorescence originates from the semiconducting band structure of protein with an optical band-gap in blue-green range of electromagnetic radiation. Delocalization of peptide electrons through peptide chain contributes significantly in determining the semiconductor band structure of proteins. The blue-green emission of protein is a result of electron-hole radiative recombination after photo-excitation. A much higher effective concentration of the band-forming peptide units are present in concentrated protein solution, crystal and also in amyloid fibril, making the net emission detectable readily. Though blue-green emission may be present in amyloid fibrils, it should not be treated as exclusive signature of amyloid fibril or aggregation process, as is practiced at present. The phenomenon may probably be common to all proteins, and protein like polymeric macromolecules. This may well be the reason for blue-green auto-fluorescence of a cell, which is packed with such molecules at very high concentration.[8] We strongly believe that our findings resolve the debate regarding the origin of blue-green fluorescence in proteins, and in other macromolecules, and will help in validating the theory of band structure of protein and similar polymers. Further advanced photophysical and photo-conductivity studies may lead to interesting and important findings regarding the relevance of (semi)conducting nature of the biomolecules and charge transfer towards the functionality of biosystems.

**ACKNOWLEDGMENTS**

SS thanks UGC, AS thanks CSIR for fellowship. We thank IISER Pune and CSIR for funding. We thank Dr. Saikrishnan Kayarat (SEC), Meghna Manae, Amod Desai, Dr. Anirban Hazra and Dr. Arnab Mukherjee for useful discussion.

**REFERENCES**

[1] A. L. Fink, Fold Des. **3**, R9 (1998).
[2] N. J. Marianayagam, M. Sunde, and J. M. Matthews, Trends Biochem. Sci. **29**, 618 (2004).
[3] D. Pinotsi *et al.*, ChemBioChem **14**, 846 (2013).
[4] U. Anand, and M. Mukherjee, Langmuir **29**, 2713 (2013).

[5]L. L. del Mercato *et al.*, *Proc. Natl. Acad. Sci. U.S.A.* **104**, 18019 (2007).

[6]N. Amdursky *et al.*, Nano Lett. **9**, 3111 (2009).

[7]A. Shukla *et al.*, *Arch. Biochem. Biophys* **428**, 144 (2004).

[8]P. Guptasarma, *Arch. Biochem. Biophys* **478**, 127 (2008).

[9]A. M. Smith *et al.*, *Adv. Mater.* **20**, 37 (2008).

[10]F. T. S. Chan *et al.*, Analyst **138**, 2156 (2013).

[11]S. Sharpe *et al.*, Biomacromolecules **12**, 1546 (2011).

[12]S. Chatterjee, and T. K. Mukherjee, J. Phys. Chem. B **117**, 16110 (2013).

[13] J. R. Lakowicz, *Principles of fluorescence spectroscopy*. Third Edition, Springer, Chapter 16 (2006).

[14]A. Samanta, J. Phys. Chem. B **110**, 13704 (2006).

[15]P. Singh, and P. K. Chowdhury, J. Phys. Chem. Lett. **4**, 2610 (2013).

[16]S. Haldar, A. Chaudhuri, and A. Chattopadhyay, J. Phys. Chem. B **115**, 13704 (2011).

[17]A. Chattopadhyay, and S. Haldar, *Acc. Chem. Res.* **47**, 12 (2014).

[18]M. Heinen *et al.*, Soft Matter **8**, 1404 (2012).

[19]B. Jachimska, M. Wasilewska, and Z. Adamczyk, Langmuir **24**, 58 (2008).

[20]R. Itri *et al.*, Braz. J. Phys. **34**, 288 (2004).

[21]I. Hayakawa *et al.*, J. Food Sci. **57** (1992).

[22]C. Kittel, *Introduction to Solid State Physics*, Eigth edition, Wiley India Edition, Chapter 8 (2005).

[23]A. Szent-Gyorgyi, Nature **148** (1941).

[24]A. Szent-Gyorgyi, Science **93** (1941).

[25]S. Baxter, *Trans. Faraday Soc.* **39**, 207 (1943).

[26]R. Pethig, J Biol Phys **6**, 1 (1978).

[27]H. B. Gray, and J. R. Winkler, *Chem. Phys. Lett.* **483**, 1 (2009).

[28]Y. Tokita *et al.*, *Angew. Chem. Int. Ed.* **50**, 11663 (2011).

[29]M. G. G. Evans, J., BBA-Mol. Basis. Dis **3**, 188 (1949).

[30]H. Fujita, and A. Imamura, J. Chem. Phys **53**, 4555 (1970).

[31]D. D. Eley, and D. I. Spivey, *Trans. Faraday Soc.* **56**, 1432 (1960).

[32]M. H. Cardew, and D. D. Eley, *Discuss. Faraday Soc.* **27**, 115 (1959).

[33]D. Wang, and T. Imae, *J. Am. Chem. Soc.* **126**, 13204 (2004).

**See supplemental material at [URL will be inserted by AIP] for** Experimental section, Trp emission (350 nm), blue emission of HSA and lysozyme, REES spectra, CD, DLS, SEC, REES in RM, effect of denaturation, blue emission in polymeric macromolecules

# Blue Emission in Proteins

Sohini Sarkar, Abhigyan Sengupta, Partha Hazra* and Pankaj Mandal*

Department of Chemistry,

Indian Institute of Science Education and Research (IISER),

Dr. Homi Bhabha Road, Pune 411008, INDIA

*To whom correspondence should be addressed:

Email: pankaj@iiserpune.ac.in, p.hazra@iiserpune.ac.in

Tel: 91-20-25908030

**Experimental Section**

Bovine Serum Albumin (BSA), Human Serum Albumin (HSA) and Lysozyme (purity ≥ 98 %) were purchased from Sigma Aldrich and used without further purification. Phosphate buffer solution at pH = 7 was used for making protein solutions and dilutions. For pH dependent study the pH of the buffer solution was varied as required. For calculating BSA, HSA and Lysozyme concentrations, their molar extinction coefficients were taken to be 43,291 $M^{-1}cm^{-1}$, 35,495 $M^{-1}cm^{-1}$, 37,646 $M^{-1}$ $cm^{-1}$ respectively at 280 nm.[1] Spectral data were recorded at experimental temperature of 298 K using freshly prepared and properly equilibrated protein solutions unless otherwise mentioned. To rule out any effect of contamination fluorescence experiments were repeated several times using different cuvette (including brand new ones).

**pH :** Silicon micro-sensor pocket-sized pH meter (ISFETCOM Co. Ltd., Japan) was used for measuring pH of buffer solutions.

**Steady State:** UV-2600 Shimadzu UV visible Spectrophotometer was used for recording absorption spectra and Fluorolog-3 Spectrofluorometer (Horiba Jobin Yvon, USA) was used for recording steady state fluorescence spectra. All spectral data were acquired using quartz cuvette of 1 cm path length. Slit width of 1 nm for was used for excitation and 5 nm for emission.

**Lifetime:** A time-correlated single photon counting (TCSPC) setup from IBH Horiba Jobin Yvon (USA) was used to collect fluorescence lifetime and time-resolved anisotropy measurements. For lifetime measurements of the fluorophore that emit around 450 nm we used a diode laser having central wavelength at 375 nm. Emission was collected over a total time range (TAC) of 60 ns at different wavelengths through single monochromator with 12 nm bandpass. IRF (instrument response function) was measured from fullwidth half maxima (FWHM) of LED scattering collected using LUDOX prompt. FWHM was in the range of 120 ps.

**CD:** Circular Dichroism (CD) spectra were recorded using J-815 CD spectropolarimeter (Jasco, USA) keeping the samples in a 1 mm pathlength quartz cuvette. After taking suitable baseline correction of the blank buffer, three scans at scan speed of 20 nm min$^{-1}$ were averaged to get a CD signal of a particular concentration of BSA.

**DLS:** Concentration dependent size determination of BSA was carried out by dynamic light scattering (DLS), employing a Nano ZS-90 apparatus that uses 633 nm red laser from Malvern instruments. Each BSA solution of a particular concentration was scanned for 3 times each scan being an average of 20 scans. Water and all the samples for DLS measurements were filtered through 0.22μm membrane filter. The entire DLS experiment for all BSA concentrations was repeated thrice to check reproducibility of the data.

**Reverse Micelle:** Dioctyl sulfosuccinate sodium salt (AOT, Sigma Aldrich, 98 % pure) was dried under high vacuum for 24 hours to eliminate the adsorbed water before use. n-Heptane (SRL, India , spectroscopic grade) was used as received. The AOT concentration was fixed at 0.2 M in the reverse micelle (RM). The $W_0$ value for water RM is defined by the molar ratio of water to surfactant, $W_0$ = [water]/[AOT ]. Water RM solution of $W_0$ = 25 was prepared by the addition of an appropriate amount of water to the AOT/ n -heptane system. Calculated amount of solid BSA was added to this water RM solution so that the effective concentration of protein in solution was around 10μM of BSA. For AOT reverse micelles, $W_0$ is directly proportional to the hydrodynamic radius ($R_H$) and $R_H$ was calculated using the equation: $R_H$(nm) = 0.175$W_0$ + 1.5.[2] For $W_0$ value of 25, the entrapped water pool within the RM had the hydrodynamic radius of 5.8 nm which can fairly accommodate single BSA protein molecule.

**Gel Filtration:** Size exclusion chromatography (SEC) of BSA was done using a Superdex 200 10/300GL column at 4°C. Two sample of pure BSA of 10 μM (blue) and 100 μM (red) concentrations were injected onto the column. (Figure S7) The column was equilibrated and the protein eluted with phosphate buffer of pH 7.

**Table S1:** Fluorescence Lifetime parameters of the emitting species at 450 nm with increasing BSA concentration measured using Time correlated single photon counting (TSCPC) set up.

| BSA(μM) | $\tau_1$ (ns) | $\alpha_1$ | $\tau_2$ (ns) | $\alpha_2$ | $\tau_3$ (ns) | $\alpha_3$ | $\tau_{avg}$ (ns) | $\chi^2$ |
|---|---|---|---|---|---|---|---|---|
| 48 | 1.4 | 0.26 | 0.14 | 0.63 | 6.82 | 0.11 | 1.19 | 1.07 |
| 240 | 1.47 | 0.25 | 0.15 | 0.65 | 6.85 | 0.1 | 1.14 | 1.09 |
| 480 | 1.56 | 0.29 | 0.18 | 0.59 | 6.89 | 0.12 | 1.37 | 1.1 |

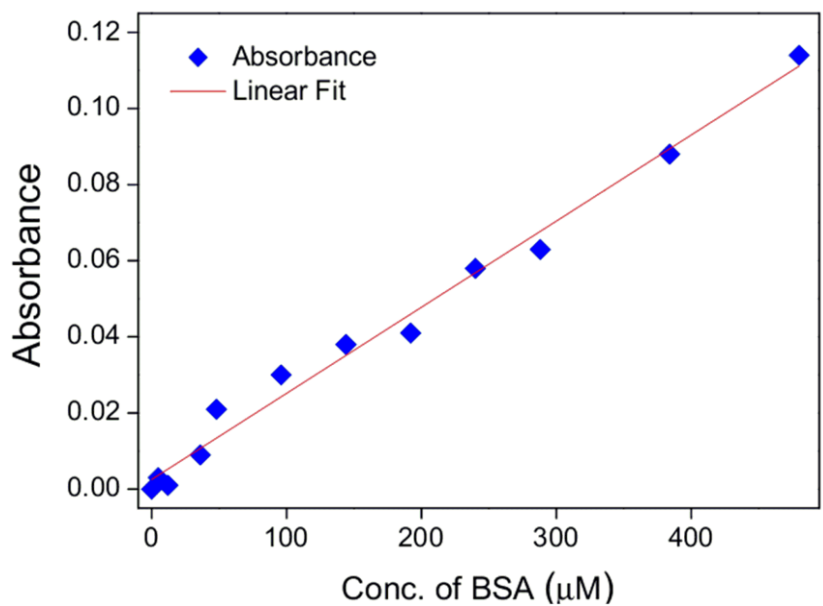

**Figure S1:** Concentration dependent absorption of BSA at 410 nm. Molar absorption coefficient value obtained from linear fit is 226 $M^{-1}$ $cm^{-1}$.

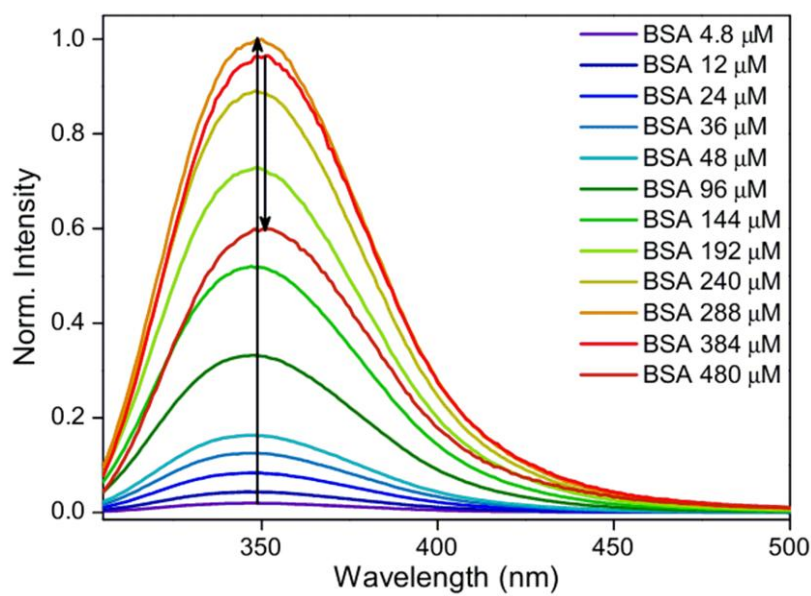

**Figure S2.** Normalized Emission spectrum of tryptophan in BSA ($\lambda_{exc}$ = 295 nm to selectively excite tryptophan). Inner filter effect was corrected for all emission intensities at each concentration using the dual path-length method.[3]

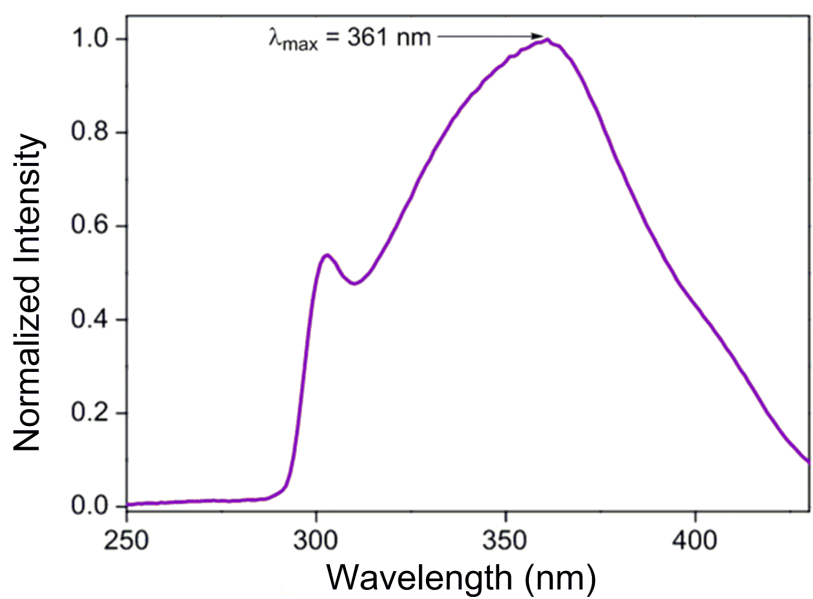

**Figure S3.** Excitation spectrum of 480 μM of BSA where emission intensity is monitored at 450 nm while scanning the excitation wavelength from 250 nm to 435 nm.

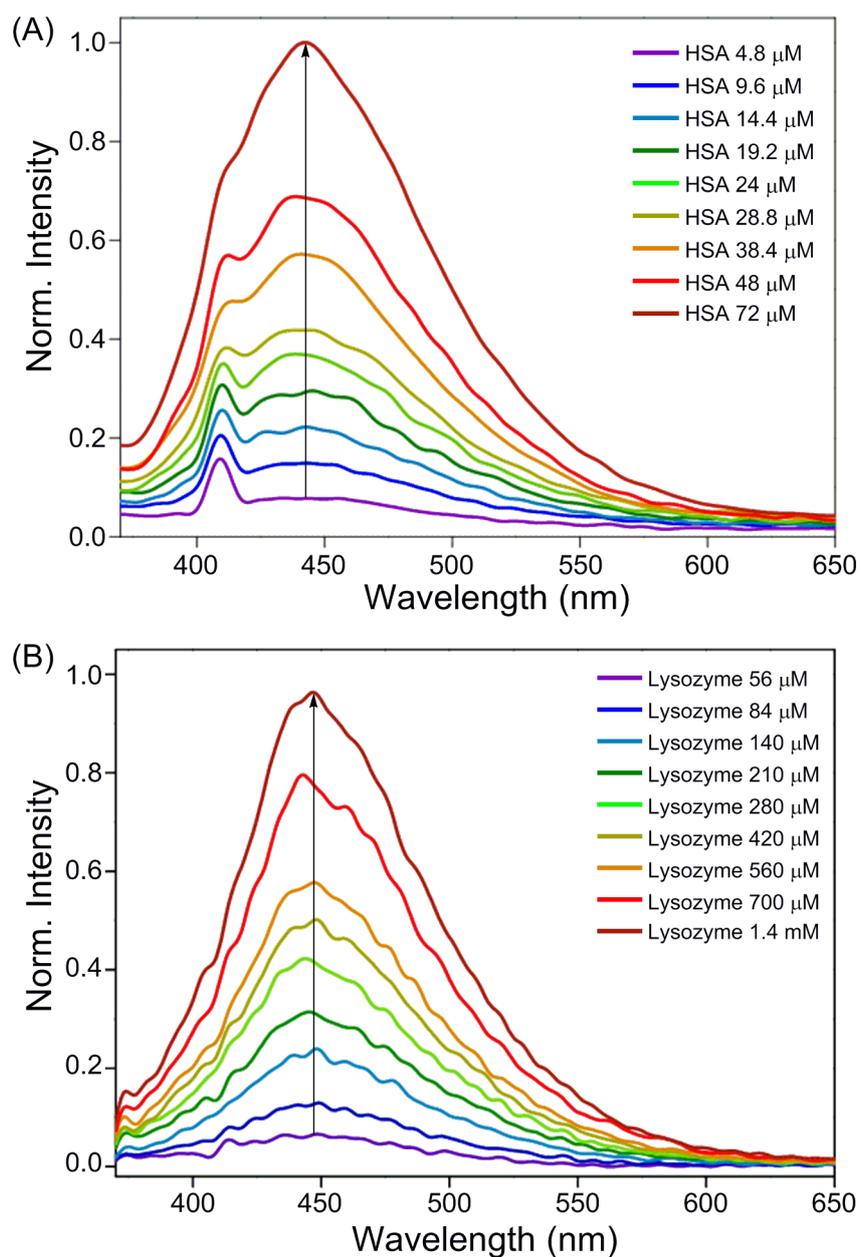

**Figure S4.** Concentration dependent normalized blue emission of (A) HSA (Mol wt. = 66.6 kD and 585 amino acid residues), (B) Lysozyme (Mol wt. = 14.3kD and 129 amino acid residues) at $\lambda_{exc}$ = 360 nm.

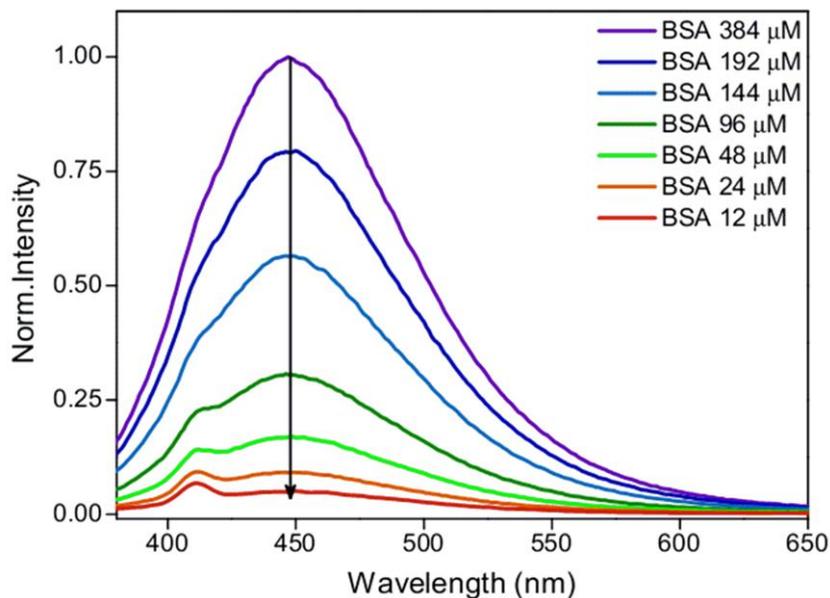

**Figure S5:** Emission profile of blue fluorescence with dilution. The emission intensities reverted to match emission profiles at initial lower concentration. This suggests the reversibility of concentration induced fluorescence.

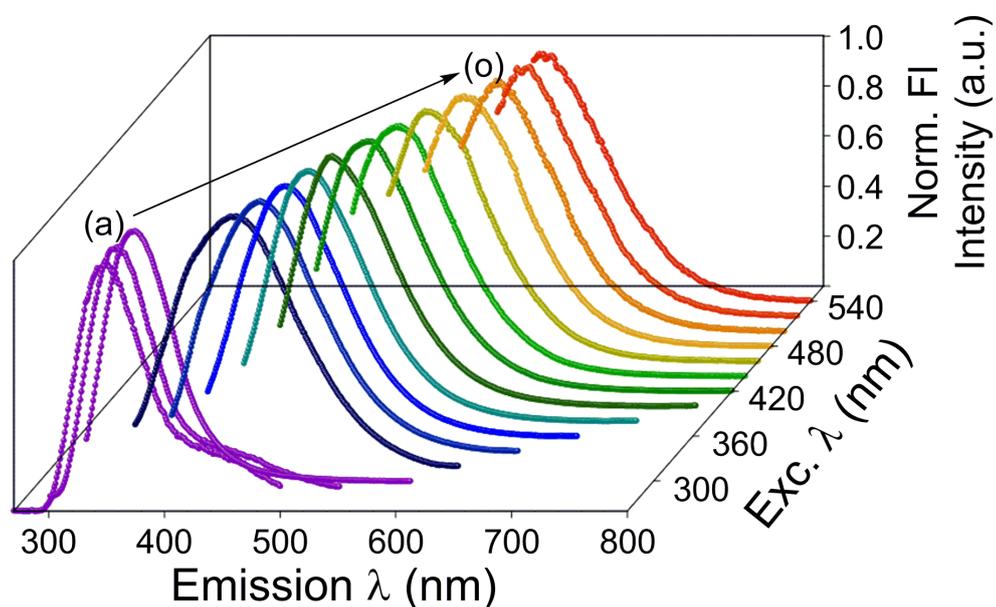

**Figure S6.** Red Edge Excitation Shift (REES) of blue emission spectra with change in excitation wavelength from 260 nm to 540 nm in a step of 20 nm. Tryptophan emission does not shift with change in excitation wavelength.

In order to understand the concentration induced secondary structure perturbation of BSA, we carried out concentration dependent circular dichroism (CD) study. At lower BSA concentration, the CD spectrum shows two minima at 209 nm and 222 nm, which are indicative of α-helical structure indigenous to BSA protein.[4] With increase in BSA concentration there was an obvious increase in helicity at 222 nm and 209 nm. However, due to very high absorption in UV at concentrations above 12μM of BSA, the detector gets saturated. So, the apparent changes in the CD spectra are merely an artefact, and nothing should be inferred from this observation.

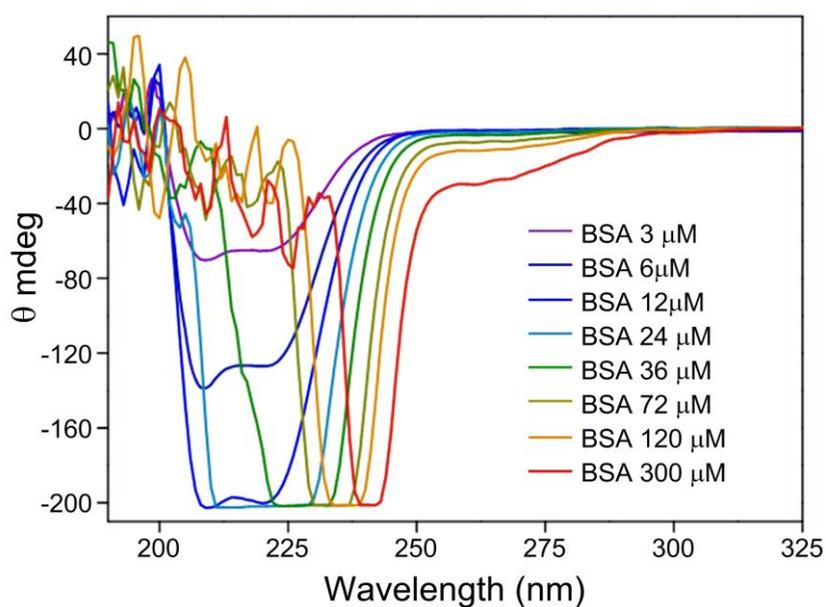

**Figure S7:** Concentration dependent CD spectra of BSA.

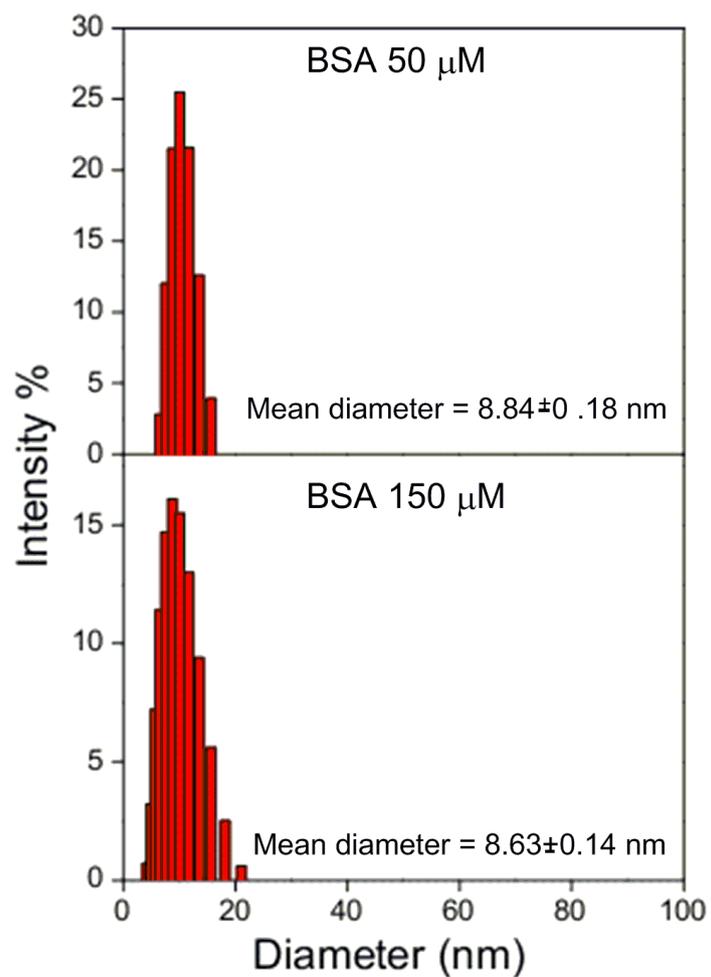

**Figure S8:** DLS Data showing mean diameter of 50 μM and 150 μM of BSA.

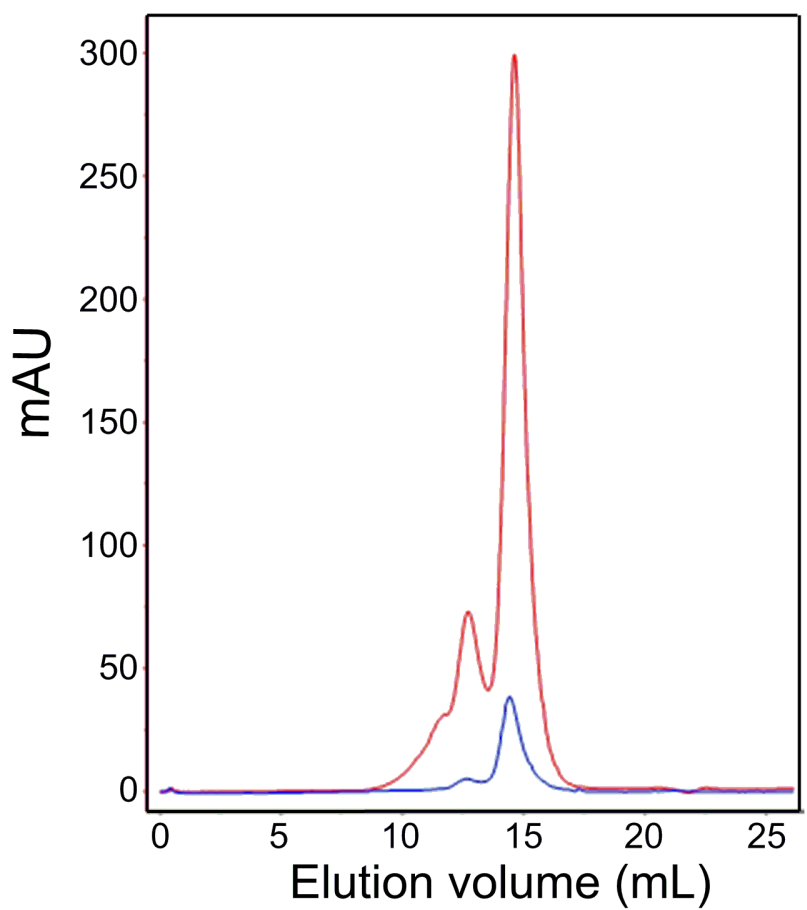

**Figure S9:** Size exclusion chromatography (SEC) of BSA using a Superdex 200 10/300GL column. Two sample of pure BSA of 100 uM (red) and 10 uM (blue) concentrations were injected onto the column. The elution volume of the major peak was about 14.6 mL, corresponding to the molecular size of monomeric BSA. A small fraction of protein eluted at a volume corresponding to dimeric and tetrameric form of BSA. However, a second round of SEC of the monomeric fraction yielded a similar profile having both monomeric and higher order structure. This indicates to an equilibrium between the monomeric and the higher order units at the experimental condition. To find out if the monomeric protein unit is capable of emitting blue light or the oligomeric structures are responsible for the observed fluorescence, we recorded fluorescence spectrum of protein molecule confined inside a reverse micelle (RM) where the size of the RM chosen excluded the possibility of having oligomers.

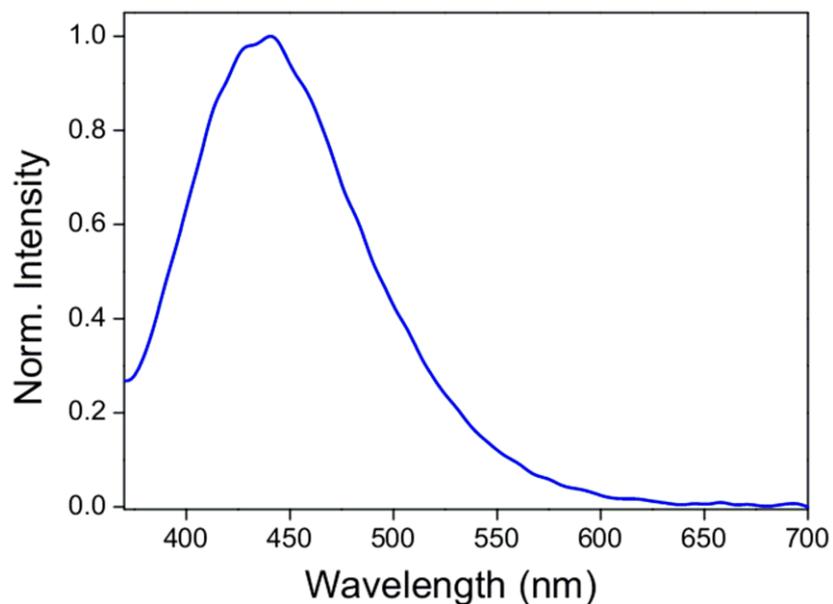

**Figure S10.** Normalized fluorescence spectrum of monomeric fraction of BSA obtained from gel filtration.

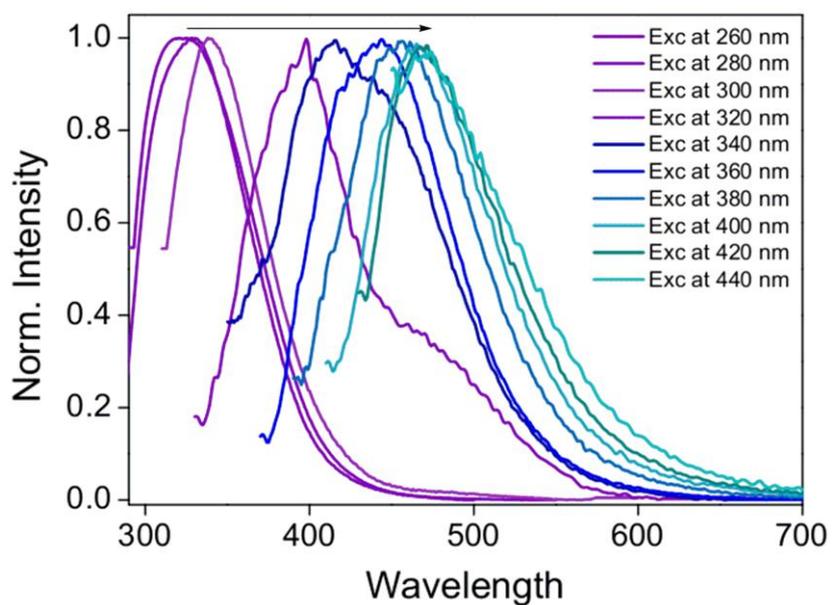

**Figure S11.** Excitation wavelength dependent emission spectra of BSA in AOT reverse micelle with change in excitation wavelength from 260 nm to 440 nm in a step of 20 nm.

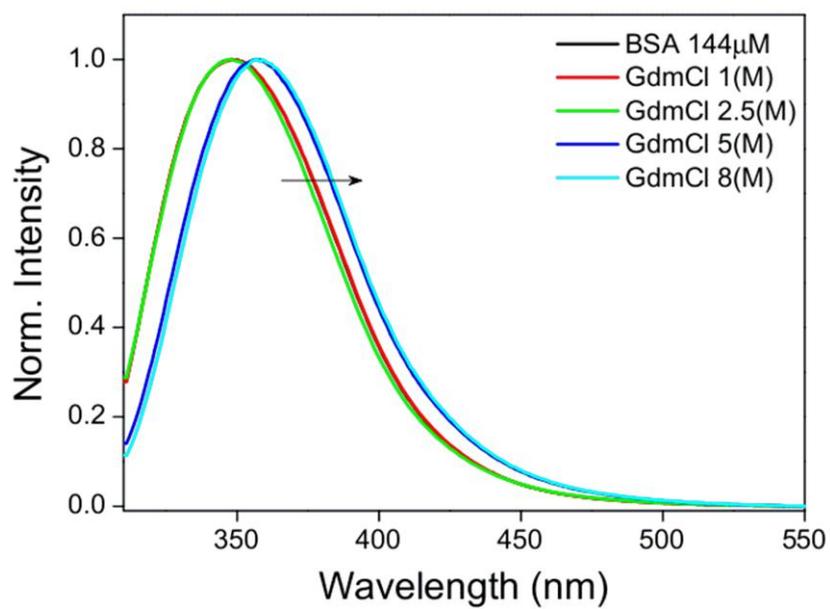

**Figure S12.** Normalized emission spectra showing red shift of tryptophan emission in 144 μM of BSA with increasing guanidinium chloride concentration.

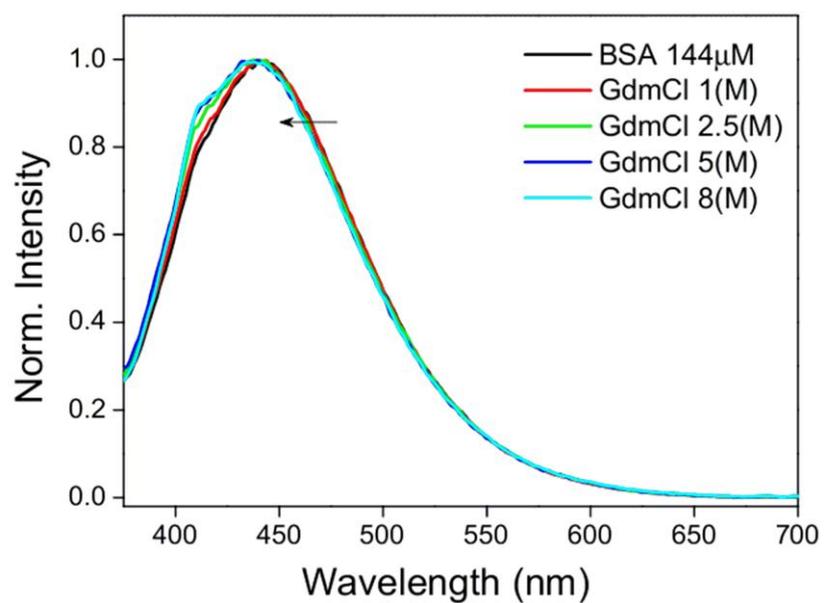

**Figure S13.** Normalized emission spectra showing blue shift of the blue emission in 144 μM of BSA with increasing guanidinium chloride concentration.

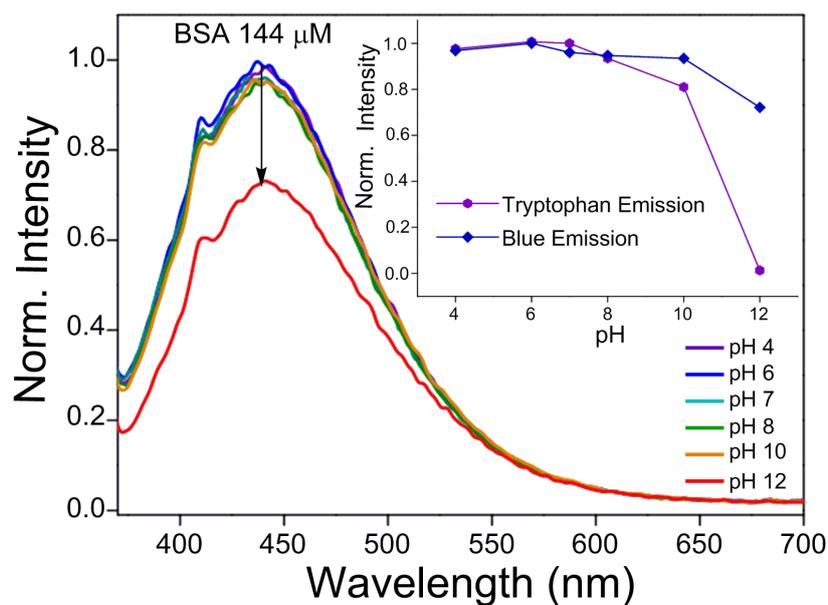

**Figure S14:** pH dependent quenching of blue fluorescence at 144 μM of BSA. Inset shows pH dependent quenching of tryptophan emission at 350 nm and blue emission at 450 nm. Only the peak emission intensities are plotted. At pH 12 there is complete denaturation of the protein.

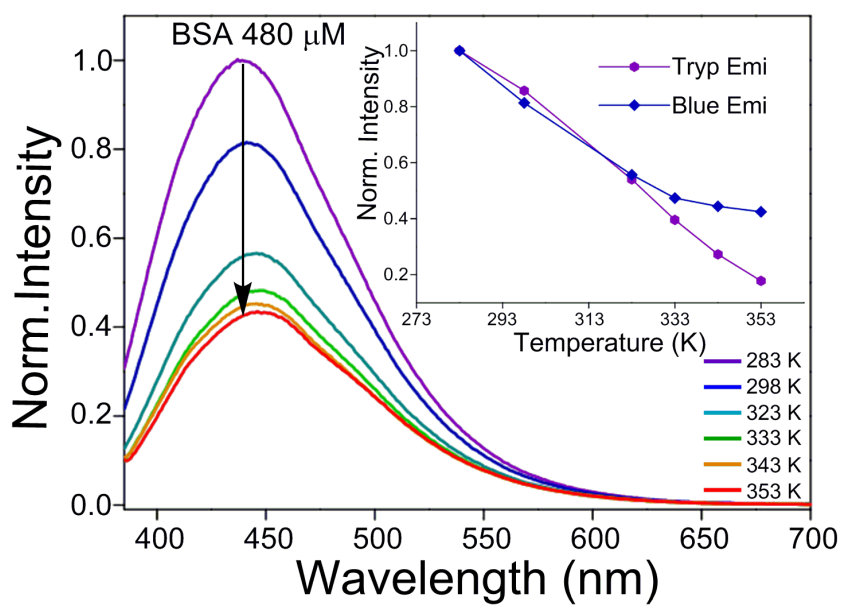

**Figure S15:** Decrease in intensity of blue fluorescence with increase in temperature. Inset shows temperature dependent quenching of tryptophan emission at 350 nm and blue emission at 450 nm. Only the peak emission intensities are plotted.

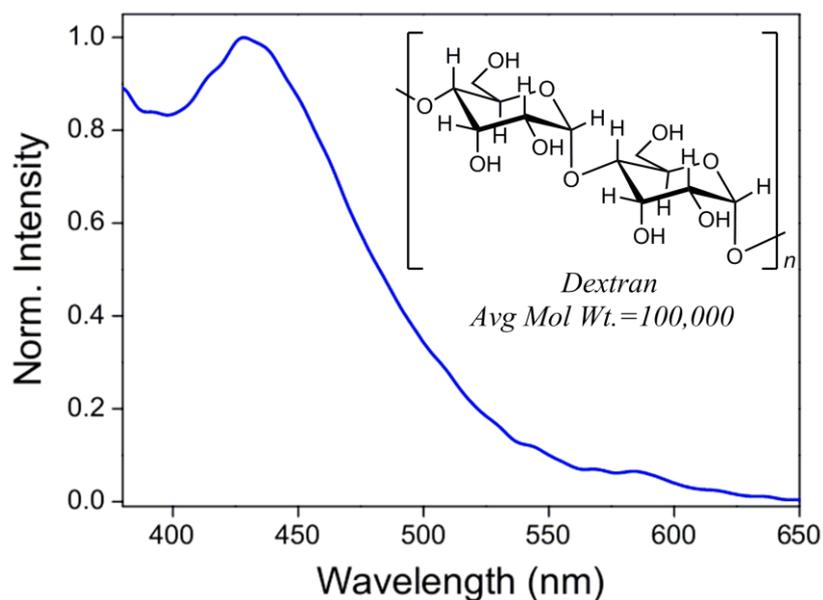

**Figure S16 (A):** Blue emission observed in Dextran ($\lambda_{exc}$ = 360 nm).

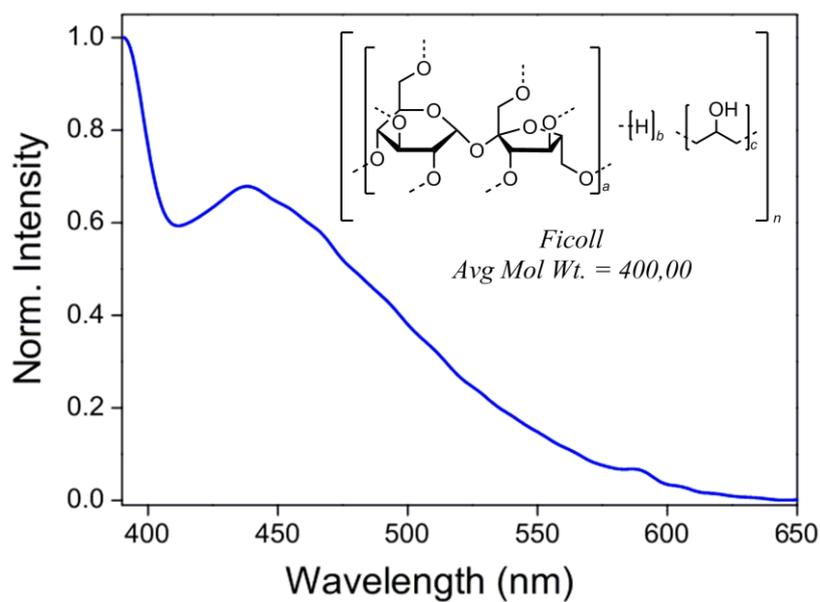

**Figure S16(B):** Blue emission observed in Ficoll ($\lambda_{exc}$ = 360 nm).

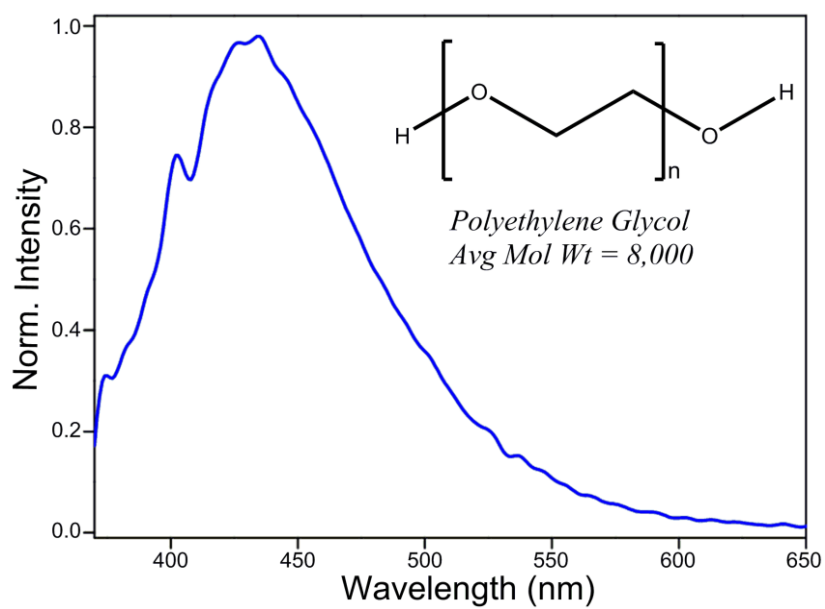

**Figure S16(C):** Blue emission observed in Polyethylene glycol ($\lambda_{exc}$ = 360 nm).

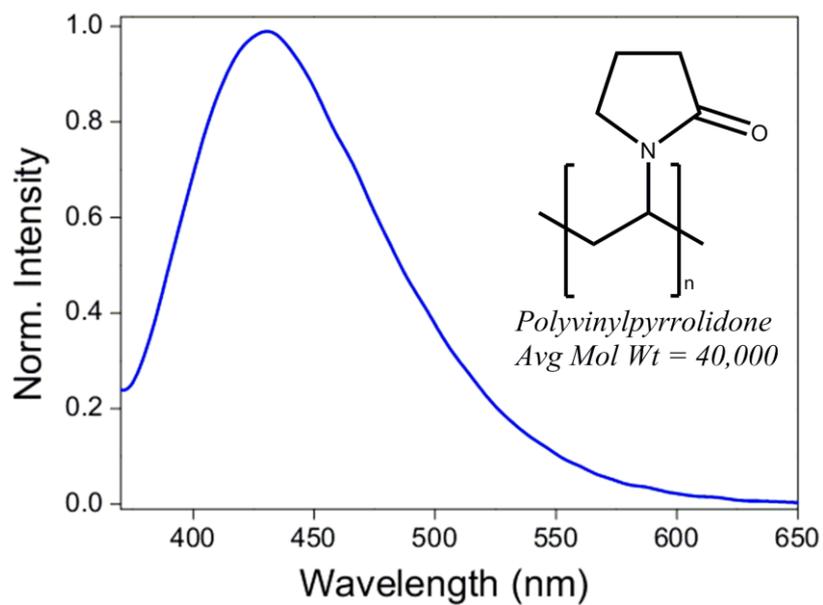

**Figure S16(D):** Blue emission observed in Polyvinylpyrrolidone ($\lambda_{exc}$ = 360 nm).

References:

1. Pace, C. N.; Vajdos, F.; Fee, L.; Grimsley, G.; Gray, T., How to measure and predict the molar absorption coefficient of a protein. *Protein Science* **1995,** *4* (11), 2411-2423.
2. McPhee, J. T.; Scott, E.; Levinger, N. E.; Van Orden, A., Cy3 in AOT Reverse Micelles I. Dimer Formation Revealed through Steady-State and Time-Resolved Spectroscopy. *The Journal of Physical Chemistry B* **2011,** *115* (31), 9576-9584.
3. Pagano, T. E.; Kenny, J. E. In *Assessment of inner filter effects in fluorescence spectroscopy using the dual-pathlength method: a study of the jet fuel JP-4*, 1999; pp 289-297.
4. Aromatic Contributions To Circular Dichroism Spectra Of Protein. *Critical Reviews in Biochemistry and Molecular Biology* **1974,** *2* (1), 113-175.